\documentclass[a4paper,11pt,noarxiv]{quantumarticle}
\pdfoutput=1
\usepackage[utf8]{inputenc}
\usepackage[T1]{fontenc}
\usepackage{amsmath,amssymb,mathtools}
\usepackage{graphicx}
\usepackage{booktabs}
\usepackage{braket}
\usepackage{nameref}
\usepackage[colorlinks=true,linkcolor=blue!60!black,citecolor=blue!60!black,urlcolor=blue!60!black]{hyperref}

\newcommand{\avqkan}{Adaptive VQKAN}
\newcommand{\Hread}{\hat H}

\title{Few-sample regression with an adaptively grown variational quantum Kolmogorov--Arnold network}
\date{\today}

\author{Hikaru Wakaura}
\email{hikaruwakaura@gmail.com}
\affiliation{QuantScape Inc., 4-11-18 Manshon-Shimizudai, Meguro, Tokyo 153-0064, Japan}
\author{Rahmat Mulyawan}
\affiliation{School of Electrical Engineering and Informatics, Institut Teknologi Bandung, Jl.\ Ganesha No.\ 10, Bandung, Indonesia}
\affiliation{Research Collaboration Center for Quantum Technology 2.0, BRIN--ITB--TelU, Indonesia}
\author{Andriyan B. Suksmono}
\affiliation{School of Electrical Engineering and Informatics, Institut Teknologi Bandung, Jl.\ Ganesha No.\ 10, Bandung, Indonesia}
\affiliation{Research Collaboration Center for Quantum Technology 2.0, BRIN--ITB--TelU, Indonesia}
\affiliation{ITB Research Center on ICT (PPTIK-ITB), Bandung, Indonesia}

\begin{document}

\maketitle

\begin{abstract}
Kolmogorov--Arnold networks place learnable one-dimensional functions on the edges of a network rather than fixed activations on its nodes, and several quantum realisations of this architecture have been proposed. Whether any of them offers a practical benefit is unclear, because the reported comparisons rest on single training runs, on baselines that were not tuned, and on test sets that were also used for model selection. Here we evaluate a variational quantum Kolmogorov--Arnold network whose ansatz is grown one Pauli operator at a time, under a protocol with seed-paired comparisons, a stopping rule that never sees the test set, and a confirmatory study whose hypotheses, seeds, and analysis were committed before the runs. On four-qubit benchmarks the model is indistinguishable from a quantum neural network of the same parameter count and is outperformed by classical regressors. On a difficulty-controlled family of 12- and 16-dimensional targets learned from ten training points, the 32-parameter quantum model beats the best of four unregularised classical regressors and a tuned quantum neural network with Holm-corrected $p\le 0.017$, the result is unchanged when every circuit is evaluated with 256 measurement shots, and the trained models run on a 156-qubit IBM processor with test errors within 0.3 of the exact values. A kernel ridge regressor whose hyperparameters are chosen by leave-one-out cross-validation on the same ten points matches the quantum model, and a sample-size sweep shows the quantum model's error is nearly flat in the training-set size while the classical models keep improving. The benefit of the quantum model is therefore an implicit regularisation of a low-capacity model, present only in the few-sample regime and absent at 18 dimensions, and not an expressivity advantage. These results give a reproducible reference point for the resources and limits of variational quantum Kolmogorov--Arnold networks.
\end{abstract}

\section{Introduction}\label{sec:intro}

Machine-learning models are increasingly used to interpolate physical quantities from a handful of expensive measurements or simulations, a regime in which the inductive bias of the model matters more than its capacity. Kolmogorov--Arnold networks (KANs)~\cite{2024arXiv240419756L} address this regime with a specific bias: the network is a sum of learnable one-dimensional functions on its edges, following the Kolmogorov--Arnold representation theorem, and the learned functions are directly interpretable. The architecture has been adopted quickly across image analysis, time-series modelling, physical-equation solving, and control~\cite{2024arXiv241106727C,2024EPJQT..11...76K}, and its structure maps naturally onto parametrised quantum circuits, where each edge function can be realised by a rotation angle that depends on a measured expectation value.

Several quantum KANs have followed. Fault-tolerant constructions implement the layer map with block encodings and quantum singular value transformation~\cite{Ivashkov2026QKAN}; near-term constructions use variational circuits~\cite{Wakaura_VQKAN_2024,Wakaura2025EVQKAN,QVAF2025}, Born-machine sampling~\cite{QuKAN2025}, or adversarial training~\cite{GAVQKAN2025}. The variational family belongs to the broader class of variational quantum algorithms~\cite{Kassal2011,McClean_2016,2020arXiv201209265C} and quantum-circuit learning models~\cite{PhysRevA.98.032309,PerezSalinas2020}, whose trainability~\cite{mcclean_barren_2018} and relation to classical kernel methods~\cite{Schuld2021kernel,Huang2021power,Jerbi2023beyond} have been studied intensively. A natural way to keep such a model small is to grow its ansatz adaptively, operator by operator, as in the ADAPT variant of the variational quantum eigensolver~\cite{2019NatCo..10.3007G}; the resulting \avqkan\ uses only a few parametrised Pauli rotations per layer.

However, no quantum KAN has yet been evaluated under conditions that would let a reader decide whether it is worth using: published comparisons rest on single optimisation runs, on quantum-neural-network baselines whose initialisation and depth were not tuned, on classical baselines without regularisation, and on test sets that were also used to pick the best epoch. In this study we evaluate the \avqkan\ under a protocol designed to remove each of these confounds. Every comparison is paired across random seeds, model selection uses only the training cost, all baselines are tuned on the training points alone, and the central claim is tested in a pre-registered study whose hypotheses, seeds, and analysis script were committed to a public repository before the runs were launched. We found that the model offers no advantage at four qubits, that it beats unregularised classical regressors and a tuned quantum neural network when ten training points are spread over 12 or 16 dimensions, and that a cross-validated kernel ridge regressor recovers the same accuracy. The results show that the useful property of the \avqkan\ is an implicit regularisation of a deliberately low-capacity model, which is genuine but classically reproducible, rather than a quantum expressivity advantage.

\section{Model and protocol}\label{sec:model}

\subsection{Adaptive variational quantum Kolmogorov--Arnold network}

The model maps an input $\mathbf{x}\in[0,1]^d$ to a scalar prediction $h(\mathbf{x};\boldsymbol{\theta})=\braket{\Psi(\mathbf{x};\boldsymbol{\theta})|\Hread|\Psi(\mathbf{x};\boldsymbol{\theta})}$ on $N_q=\max(4,d)$ qubits. The input is encoded by one rotation per coordinate, $R_y\!\left(\arccos x_j+\pi/2\right)$ on qubit $j<d$, the remaining qubits starting in $\ket{0}$. A KAN layer then consists of an ordered list of $N_p$ Pauli strings $P_p$, each carrying one learnable edge function. The layer first measures the single-qubit expectations $z_i=\tfrac12(\braket{Z_i}+1)$ of the current state, and then applies
\begin{align}
U(\mathbf{z};\boldsymbol{\theta})&=\prod_{p=1}^{N_p}\exp\!\big(-i\pi\,\phi_p(\mathbf{z})\,P_p\big),\nonumber\\
\phi_p(\mathbf{z})&=\Pi\!\left[\sum_{i=0}^{d-1}2\arccos\!\frac{\sigma_p(z_i)+E(z_i)-\sigma_p^{\min}}{\sigma_p^{\max}+E(1)-\sigma_p^{\min}}\right],
\label{eq:phi}
\end{align}
where $\sigma_p$ is a quadratic B-spline with $N_g=8$ learnable control values on a uniform grid over $[-1,1]$, evaluated at the input value, $\sigma_p^{\min/\max}$ are its extrema over the grid, $E(z)=z/(1+e^{-z})$ is a fixed activation, the argument of the arccosine is clipped to $[-1,1]$, and $\Pi[\cdot]$ maps the summed angle modulo $2\pi$ onto $[-1,1]$. Each operator thus contributes $N_g$ trainable parameters, and a model with $N_p$ operators has $8N_p$ parameters in total. The readout is $\Hread=Z_0Z_1$ for the four-qubit benchmarks and $\Hread=(2/N_q)\sum_j Z_{2j}Z_{2j+1}$ for the scaling family, so that $h\in[-1,1]$ in both cases. We use a single KAN layer throughout; a two-layer variant is reported in Sec.~\ref{sec:ablation}.

The ansatz is grown adaptively. Training starts from a single operator $X_1$ and, every $T_g$ optimisation steps, every operator of a pool---all one-body strings $X_j,Y_j,Z_j$ and the two-body strings $XX,XY,XZ,YY,YZ,ZZ$ on every pair---is appended tentatively with its control values set to zero; the operator that lowers the training cost most is adopted, and no operator is adopted if none lowers it. This is the selection rule of ADAPT-VQE~\cite{2019NatCo..10.3007G} with the cost in place of the energy. A gradient-based selection rule and a larger pool are compared in Sec.~\ref{sec:ablation}.

\subsection{Cost, optimiser, and budget}

Given $N$ training pairs $(\mathbf{x}_m,f_m)$ the cost is a weighted squared error,
\begin{align}
C(\boldsymbol{\theta})&=\sum_{m=1}^{N}a_m\big(h(\mathbf{x}_m;\boldsymbol{\theta})-f_m\big)^2,\nonumber\\
a_m&=\frac{B-(f_m-\min_{m'}f_{m'})}{B},\qquad B=2+\frac1N,
\label{eq:cost}
\end{align}
which weights samples with small target values slightly more. The parameters are updated by the derivative-free COBYLA method~\cite{Powell1994} in a sequence of $S$ steps, each step restarting COBYLA from the current parameters with an iteration cap $M$; the ansatz grows every $T_g$ steps. We use $S=60$, $T_g=20$, $M=200$ for the four-qubit benchmarks and $S=40$, $T_g=10$, $M=30$ for the scaling family. The number of circuit evaluations consumed by the optimiser is counted exactly and reported in Table~\ref{tab:resources}. A companion study compares this update rule with natural-gradient and plain gradient descent on the same model~\cite{Wakaura2025NG}; here the model, not the optimiser, is the object of study.

\subsection{Targets}

The four-qubit benchmarks are five synthetic regressions with two to four input variables, all rescaled to $[-1,1]$: the main target
\begin{equation}
f(\mathbf{x})=2\,\frac{\exp\!\big(\sin(x_0^2+x_1^2)+\sin(x_2^2+x_3^2)\big)-1}{e^{2}-1}-1,
\label{eq:eq6}
\end{equation}
a logarithmic target $\log(x_0/x_1)$, a fractional target $1/(1+x_0x_1)$, the radius $\sqrt{x_0^2+x_1^2+x_2^2}$ of a sphere, and an exponential target $\exp\!\big((x_0-x_1)^2/(2x_2+0.2)\big)$, with the exact normalisations given in the public code. For the scaling study we use a family whose difficulty is controlled across dimension,
\begin{equation}
\tilde f_d(\mathbf{x})=\mathrm{clip}_{[-1,1]}\!\left[\frac{\sum_{i<d/2}\sin\!\big(x_{2i}^2+x_{2i+1}^2\big)-\mu_d}{2.5\,\sigma_d}\right],
\label{eq:sepd}
\end{equation}
where $\mu_d$ and $\sigma_d$ are the mean and standard deviation of the sum over uniform inputs, fixed once by Monte Carlo and stored in the code. The standardisation makes the location and scale of the target independent of $d$, so that the error of a constant predictor is nearly the same at every size (17--18 on our metric) and results at different $d$ can be compared.

\subsection{Baselines}

\textbf{Quantum neural network.} A hardware-efficient circuit of $L$ layers, each consisting of $R_y$ rotations on all qubits, a ladder of $CZ$ gates, data-encoding rotations $R_x(2x_j)$ on the first $d$ qubits, a second $R_y$ layer, and a second $CZ$ ladder, with $2N_qL$ parameters and the same readout and cost as the \avqkan. We report two arms: a three-layer circuit with 24 parameters trained by COBYLA from a single random initialisation, which is the configuration used in published quantum-KAN comparisons~\cite{Wakaura_VQKAN_2024,Wakaura2025EVQKAN}, and a tuned arm in which $L\in\{1,2,3,4\}$ is trained by L-BFGS-B from five random initialisations and the depth is selected per seed by the training cost alone.

\textbf{Classical regressors.} A one-layer classical KAN~\cite{2024arXiv240419756L} with eight B-spline control points per input, which is linear in its coefficients and solved in closed form; multilayer perceptrons with hidden layers $8\times8$ and $16$ (tanh, L-BFGS); linear regression; and constant predictors. Their per-seed minimum, which uses test knowledge and therefore favours the classical side, is the \emph{oracle} arm of the pre-registered study. In addition we report regularised regressors whose hyperparameters are chosen on the training points only: ridge regression and the classical KAN with the ridge penalty chosen by leave-one-out cross-validation over $\lambda\in[10^{-4},10^{2}]$, kernel ridge regression with a radial basis function kernel and its penalty and width chosen by leave-one-out cross-validation on a $6\times7$ grid, and Gaussian-process regression with an anisotropic RBF kernel and white noise fitted by marginal likelihood~\cite{Rasmussen2006}. All classical outputs are clipped to $[-1,1]$.

\subsection{Statistical protocol and pre-registration}\label{sec:protocol}

Every configuration is run with a fixed set of random seeds; a seed fixes the $N=10$ training points and the 50 test points, so that all comparisons between methods are paired. The reported error is the sum of absolute deviations over the 50 test points, evaluated \emph{once}, at the optimisation step of minimum training cost; the test set is never used for any selection. Paired comparisons use the two-sided Wilcoxon signed-rank test. Dispersions are sample standard deviations.

The scaling claim was tested in a pre-registered confirmatory study. Before any run was launched, a document specifying the four hypotheses, the seeds (30--49, disjoint from all seeds used during development), the protocol, two validity gates, and the frozen analysis script was committed to the public repository (commit \texttt{74a528d}). The hypotheses were that the \avqkan\ has lower test error than the classical oracle at $d=12$, $16$, and $18$ in exact simulation and at $d=12$ with 256 measurement shots per circuit evaluation, tested with Holm correction over the family of four at $\alpha=0.05$; a pass requires the adjusted $p<0.05$ and a negative median paired difference. The validity gates require both arms to beat the constant predictor in mean. All other comparisons in this paper are exploratory and are reported with Benjamini--Hochberg $q$-values where several are made together. The comparison with the cross-validated kernel ridge regressor was added after the development seeds had been analysed and is labelled as such.

Simulations use the blueqat state-vector simulator~\cite{Kato}; finite-shot runs sample every expectation value seen by the optimiser from the Born distribution, while the reported error is always exact. For gate noise and hardware execution the trained circuits are translated to Qiskit (the translation is verified to reproduce the blueqat state vector to $10^{-8}$) and compiled to the native gate set $\{CZ, R_z, \sqrt{X}, X\}$. Because the encoding prepares a product state, the layer inputs $z_i=\tfrac12(1-\sqrt{1-x_i^2})$ are known in closed form, and inference needs one circuit per input: the encoding rotations, the adopted Pauli rotations, and a $Z$-basis measurement. Gate noise is simulated with Qiskit Aer by density-matrix evolution with depolarising channels of strength $p_2$ on two-qubit gates and $p_2/10$ on single-qubit gates, and with the calibrated noise model of the device. Hardware runs use the 156-qubit IBM Heron processor \texttt{ibm\_marrakesh} through the Sampler primitive with 2048 shots per circuit, dynamical decoupling, and measurement twirling, without readout-error mitigation.

\section{Results}\label{sec:results}

\subsection{Four-qubit benchmarks}

\begin{figure*}[t]
\centering
\includegraphics[width=\textwidth]{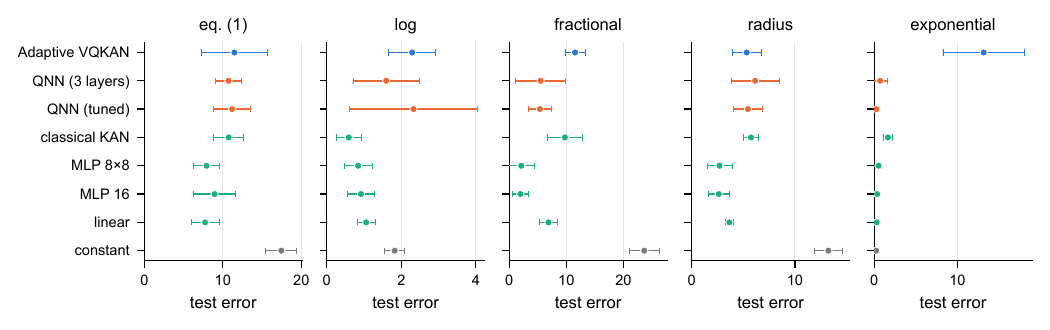}
\caption{\textbf{At four qubits the \avqkan\ is matched by a quantum neural network of equal size and beaten by classical regressors.} Test error (sum of absolute deviations over 50 test points, mean $\pm$ s.d.\ over 10 paired seeds, evaluated at the step of minimum training cost) on the five four-qubit targets. Blue: \avqkan\ (22--24 parameters). Orange: quantum neural network with three layers and 24 parameters, and with depth and initialisation tuned on the training cost. Green: classical KAN, multilayer perceptrons, and linear regression trained on the same ten points. Grey: constant predictor.}
\label{fig:d4}
\end{figure*}

Figure~\ref{fig:d4} compares all models on the five four-qubit targets. On the main target of Eq.~\eqref{eq:eq6} the \avqkan\ reaches $11.5\pm4.2$ with 24 parameters, the three-layer quantum neural network $10.7\pm1.7$ with the same 24 parameters, and the tuned network $11.2\pm2.4$; none of the three differ significantly ($p\ge0.77$). Linear regression with five parameters reaches $7.7\pm1.8$ and a multilayer perceptron $7.9\pm1.6$, both significantly better than the quantum model ($q\le0.017$). The picture is the same on the other targets: the quantum neural network is never worse than the \avqkan\ and is significantly better on the fractional and exponential targets ($q\le0.009$), and on the exponential target the quantum model ($13.2\pm4.9$) is worse than the constant predictor ($0.24\pm0.13$), whose target distribution is concentrated near $-1$. Four qubits with ten training points therefore provide no evidence of a benefit from the Kolmogorov--Arnold structure, and a single-run comparison against a randomly initialised quantum neural network can produce a gap in either direction by chance: the ten-seed mean of the three-layer network on Eq.~\eqref{eq:eq6} is 10.7, while individual seeds range from 8.4 to 13.5.

\subsection{Pre-registered scaling study}

\begin{figure}[t]
\centering
\includegraphics[width=\columnwidth]{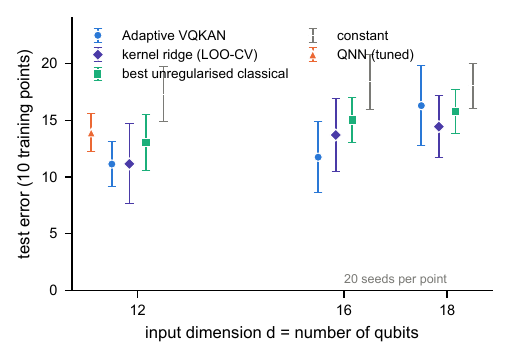}
\caption{\textbf{With ten training points in 12 and 16 dimensions the \avqkan\ beats unregularised classical regressors and a tuned quantum neural network, matches a cross-validated kernel ridge regressor, and loses its advantage at 18 dimensions.} Test error (mean $\pm$ s.d., 20 pre-registered seeds per size) on the standardised family of Eq.~\eqref{eq:sepd}. Blue circles: \avqkan\ (32 parameters). Violet diamonds: kernel ridge regression with penalty and width chosen by leave-one-out cross-validation on the ten training points. Green squares: per-seed best of classical KAN, two multilayer perceptrons, and linear regression. Grey bars: constant predictor. Orange triangle: quantum neural network with depth tuned on the training cost. The three exact-simulation hypotheses of the pre-registered study pass at $d=12$ and $16$ (Holm-adjusted $p=0.017$ and $0.004$) and fail at $d=18$.}
\label{fig:scaling}
\end{figure}

Figure~\ref{fig:scaling} shows the confirmatory study. Both validity gates hold at every size. At $d=12$ the \avqkan\ reaches $11.1\pm2.0$ against $13.0\pm2.5$ for the classical oracle (median paired difference $-2.4$, Holm-adjusted $p=0.017$), and at $d=16$ it reaches $11.7\pm3.1$ against $15.0\pm2.0$ (median difference $-4.0$, adjusted $p=0.004$). Against the individual classical models the margins are larger: at $d=16$ the classical KAN reaches $15.1\pm1.9$, the perceptrons $19.6\pm3.6$ and $20.5\pm3.5$, and linear regression $21.3\pm2.9$, all worse than the constant predictor. The tuned quantum neural network at $d=12$ reaches $13.9\pm1.7$ and is beaten by the \avqkan\ on every one of the 20 seeds ($p<10^{-4}$). At $d=18$ the quantum model reaches $16.3\pm3.5$, indistinguishable from the classical oracle ($15.8\pm1.9$, $p=1.0$) and only slightly below the constant predictor ($18.0\pm2.0$); this hypothesis fails.

The cross-validated kernel ridge regressor, added after the development seeds had been analysed, changes the interpretation of the two passing hypotheses. On the same 20 seeds it reaches $11.2\pm3.5$ at $d=12$ ($p=0.96$ against the quantum model), $13.7\pm3.2$ at $d=16$ ($p=0.048$, the quantum model slightly better), and $14.4\pm2.7$ at $d=18$ ($p=0.23$). Ridge regression and the cross-validated classical KAN do not recover the same accuracy ($17.2$ and $15.0$ at $d=12$), nor does the Gaussian process ($17.0$), whose marginal-likelihood fit of 13 kernel hyperparameters from ten points is itself under-determined. The quantum model is therefore matched, at the sizes where it beats the unregularised baselines, by a classical kernel method with two hyperparameters chosen by cross-validation on the same ten points.

\subsection{Robustness to measurement shots, gate noise, and hardware execution}

\begin{figure*}[t]
\centering
\includegraphics[width=\textwidth]{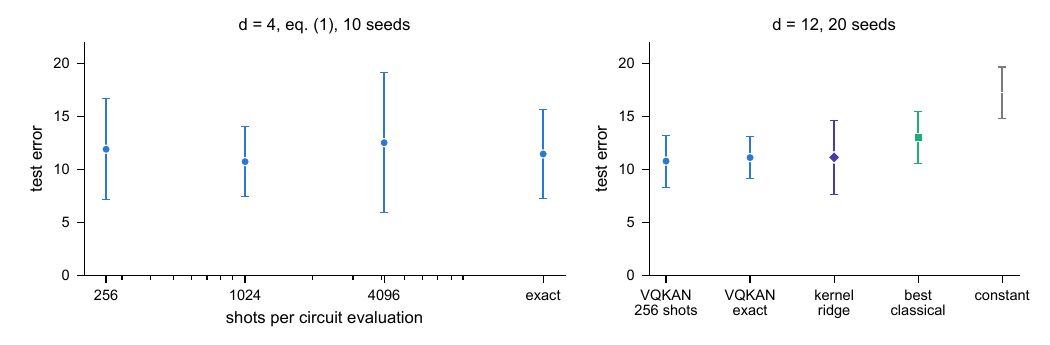}
\caption{\textbf{Finite measurement shots do not degrade the \avqkan.} Left: test error on Eq.~\eqref{eq:eq6} at four qubits when every expectation value seen by the optimiser is estimated from 256, 1024, or 4096 shots, compared with exact simulation (10 seeds; no difference is significant, $p\ge0.49$). Right: the pre-registered 256-shot arm at $d=12$ (20 seeds) next to the exact arm and the classical references of Fig.~\ref{fig:scaling}. The reported error is always computed exactly.}
\label{fig:shots}
\end{figure*}

Figure~\ref{fig:shots} shows the effect of finite measurement statistics. At four qubits the test error on Eq.~\eqref{eq:eq6} is $11.9\pm4.8$ with 256 shots, $10.8\pm3.3$ with 1024, and $12.5\pm6.6$ with 4096 shots, against $11.5\pm4.2$ in exact simulation; no difference is significant ($p\ge0.49$). At $d=12$ the pre-registered 256-shot hypothesis passes: the quantum model reaches $10.8\pm2.5$, better than the classical oracle by a median $3.0$ (adjusted $p=0.013$) and still indistinguishable from the kernel ridge regressor ($p=0.57$). Because COBYLA uses only cost values, the shot noise enters as a perturbation of the objective that the linear-interpolation model of the optimiser averages over, and the model's low capacity leaves little room for noise-driven overfitting. Shot noise is therefore not the obstacle to running this model on hardware; the number of circuit evaluations is (Table~\ref{tab:resources}).

Gate noise is equally benign, because the trained circuits are shallow: after compilation a four-qubit model contains two to four $CZ$ gates and a twelve-qubit model four to six, at depths of 13 to 21 (Table~\ref{tab:noise}). Under depolarising noise the test error of the trained four-qubit models does not change up to $p_2=0.03$, a two-qubit error rate an order of magnitude above current devices, and the twelve-qubit models behave the same way ($11.1\pm2.0$ at $p_2=0$ and $11.2\pm2.1$ at $p_2=0.03$, $p\ge0.10$). Training under noise costs a little: ten four-qubit models trained from scratch with $p_2=0.01$ and 1024 shots on every cost evaluation reach $13.1\pm4.4$, against $11.5\pm4.2$ for the same seeds trained noise-free (paired $p=0.037$), a degradation of 1.6 that is small compared with the seed-to-seed spread. Finally, we executed the trained models on the IBM Heron processor \texttt{ibm\_marrakesh}: for six four-qubit and five twelve-qubit models the test error measured on hardware differs from the exact value by at most 0.28; the mean difference over the eleven models is $-0.10\pm0.12$, a systematic reduction of about one per cent ($p=0.032$) consistent with depolarisation contracting the readout expectation toward zero, and the twelve-qubit runs are, to our knowledge, the first hardware evaluation of a quantum Kolmogorov--Arnold network at this size. The whole hardware study consumed 337 seconds of processor time.

\begin{table}[t]
\centering
\caption{Gate noise and hardware execution of the trained models (test error at the exact parameters; mean $\pm$ s.d.\ over 10 four-qubit and 20 twelve-qubit models). Depolarising strength $p_2$ on two-qubit gates, $p_2/10$ on single-qubit gates, 4096 shots. Hardware row: \texttt{ibm\_marrakesh}, 2048 shots, mean over the six four-qubit (seeds 0--5) and five twelve-qubit (seeds 30--34) models that were run, with the exact value of the same subset in parentheses.}
\label{tab:noise}
\footnotesize
\setlength{\tabcolsep}{4pt}
\resizebox{\columnwidth}{!}{%
\begin{tabular}{lcc}
\toprule
condition & $d=4$ & $d=12$ \\
\midrule
$CZ$ gates per circuit & 2--4 & 4--6 \\
exact & $11.5\pm4.2$ & $11.1\pm2.0$ \\
$p_2=10^{-3}$ & $11.5\pm4.2$ & $11.1\pm2.0$ \\
$p_2=10^{-2}$ & $11.4\pm4.1$ & $11.2\pm2.0$ \\
$p_2=3\times10^{-2}$ & $11.3\pm3.9$ & $11.2\pm2.1$ \\
device noise model & $11.4\pm4.1$ & $11.2\pm2.0$ \\
hardware (exact, same subset) & $11.5$ (exact $11.7$) & $11.1$ (exact $11.1$) \\
\bottomrule
\end{tabular}}
\end{table}

\subsection{Mechanism: sample size and ansatz composition}

\begin{figure*}[t]
\centering
\includegraphics[width=\textwidth]{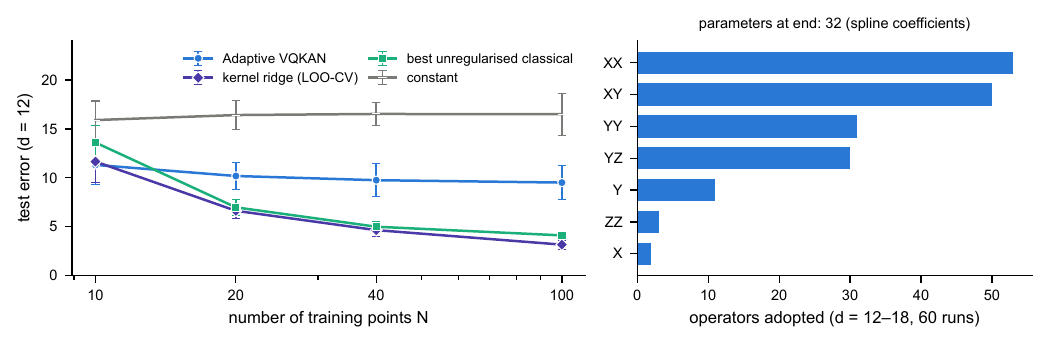}
\caption{\textbf{The advantage of the \avqkan\ is confined to the few-sample regime and is a capacity effect.} Left: test error at $d=12$ (10 seeds) as the number of training points grows from 10 to 100. The quantum model improves from $11.3$ to $9.5$ only, while the cross-validated kernel ridge regressor and the best unregularised classical model fall to $3.2$ and $4.1$. Right: operators adopted by the growth rule in the 60 pre-registered runs at $d=12$--$18$; two-body $XX$ and $XY$ strings dominate, and every run ends with four operators and 32 parameters.}
\label{fig:nsweep}
\end{figure*}

The sample-size sweep in Fig.~\ref{fig:nsweep} identifies the origin of the few-sample result. With ten training points at $d=12$ the quantum model ($11.3\pm2.1$) beats the classical oracle ($13.6\pm1.8$, $p=0.006$) and ties the kernel ridge regressor ($11.7\pm2.2$). With 20 points the order reverses ($10.2\pm1.4$ against $7.0\pm0.8$ and $6.6\pm0.8$, $p=0.002$), and with 100 points the quantum model has improved only to $9.5\pm1.7$ while the kernel ridge regressor reaches $3.2\pm0.5$. The quantum model's error is nearly flat in the training-set size: its 32 parameters, entering through eight-point splines and four Pauli rotations, cannot represent the target family beyond a fixed accuracy, and this same low capacity is what prevents it from overfitting ten points in twelve dimensions. Unregularised classical models with 13 to 200 parameters interpolate the ten points and generalise poorly; a kernel method with a cross-validated width and penalty is regularised to the same effect. The right panel shows what the growth rule selects: of the 180 operators adopted across the 60 pre-registered runs, 53 are $XX$, 50 are $XY$, 31 are $YY$, and 30 are $YZ$ strings, with one-body operators adopted 13 times. Every run adopts an operator at each of the three growth events, so the pre-registered models all end with four operators and 32 parameters.

\subsection{Growth rule, pool, and initial ansatz}\label{sec:ablation}

Table~\ref{tab:ablation} varies the components of the growth procedure on Eq.~\eqref{eq:eq6} at four qubits. Selecting the operator with the largest cost gradient instead of the largest cost decrease gives $9.3\pm2.7$ against $11.5\pm4.2$ for the cost-based rule, a nominal improvement ($p=0.037$) that does not survive correction across the six comparisons ($q=0.11$). Extending the pool to all nine two-body Pauli strings per pair changes nothing, because the additional strings are never adopted. Starting from $Y_1$ or $Z_1$ instead of $X_1$, or using two KAN layers, does not improve the result. The growth procedure is thus robust to these choices and is not the source of the model's behaviour.

\begin{table}[t]
\centering
\caption{Ablation of the growth procedure on Eq.~\eqref{eq:eq6} at four qubits (10 paired seeds, test error at the step of minimum training cost). $p$ is the paired Wilcoxon test against the first row; $q$ the Benjamini--Hochberg value over the six comparisons.}
\label{tab:ablation}
\small
\resizebox{\columnwidth}{!}{%
\begin{tabular}{lccc}
\toprule
configuration & test error & $p$ & $q$ \\
\midrule
cost selection, 6-string pool, $X_1$ & $11.5\pm4.2$ & -- & -- \\
gradient selection & $9.3\pm2.7$ & 0.037 & 0.11 \\
9-string pool & $11.5\pm4.2$ & 0.75 & 0.77 \\
gradient selection, 9-string pool & $9.2\pm2.7$ & 0.037 & 0.11 \\
start $Y_1$ & $14.6\pm4.9$ & 0.13 & 0.20 \\
start $Z_1$ & $12.2\pm3.4$ & 0.77 & 0.77 \\
two KAN layers & $15.6\pm3.6$ & 0.084 & 0.17 \\
\bottomrule
\end{tabular}}
\end{table}

\begin{table}[t]
\centering
\caption{Resources. Trainable parameters, parametrised Pauli operators at the end of training, and circuit evaluations consumed by the optimiser per training run (means over seeds; statevector preparations, excluding those used to record metrics). The quantum neural network uses 24 rotation angles in a fixed circuit.}
\label{tab:resources}
\footnotesize
\setlength{\tabcolsep}{4pt}
\resizebox{\columnwidth}{!}{%
\begin{tabular}{lccc}
\toprule
model & parameters & operators & circuit evaluations \\
\midrule
\avqkan, $d=4$, Eq.~(\ref{eq:eq6}) & 24 & 3 & 109{,}000 \\
QNN, $d=4$ & 24 & -- & 10{,}000 \\
\avqkan, $d=12$ & 32 & 4 & 25{,}400 \\
\avqkan, $d=16$ & 32 & 4 & 35{,}500 \\
\bottomrule
\end{tabular}}
\end{table}

\section{Discussion}\label{sec:discussion}

The pre-registered study establishes a real and reproducible property of the \avqkan: with ten training points in 12 or 16 dimensions, a model with 32 parameters and four Pauli rotations generalises better than a classical KAN, two perceptrons, linear regression, and a tuned quantum neural network, in exact simulation and with 256 shots. The exploratory results establish equally clearly what this property is not. It is not present at four qubits, it disappears once 20 training points are available, it disappears at 18 dimensions, and it is matched by kernel ridge regression with two hyperparameters chosen by cross-validation. The property is the implicit regularisation of a model whose capacity is bounded by construction, and the correct classical comparator for such a model is a regularised one.

This conclusion is consistent with the view that variational quantum models trained on expectation values are kernel methods in a feature space fixed by the encoding~\cite{Schuld2021kernel}, and that the relevant question for any such model is whether the geometry of its feature space suits the data better than that of an accessible classical kernel~\cite{Huang2021power,Jerbi2023beyond}. Our sample-size sweep answers this question negatively for the targets studied here: the quantum model's error saturates near 9.5 while the RBF kernel keeps improving, so the quantum feature space is not richer, only more constrained. Whether a different encoding or readout, or a KAN structure applied to targets that are themselves sums of one-dimensional functions in a rotated basis, changes this answer is open, and the protocol and code released here make such a test a matter of running the same scripts on a new target.

The failure at $d=18$ marks the capacity ceiling of the present construction. With ten points the training error of the quantum model still falls to between one and two, so the model interpolates the data, but its test error rises to the level of the constant predictor: at this dimension the four adopted operators and 32 spline coefficients no longer constrain the function enough to generalise. A larger grid, more operators, or a second layer would raise the capacity, but Table~\ref{tab:ablation} shows that at four qubits the second layer already hurts; how capacity should scale with dimension is the design question that the standardised family of Eq.~\eqref{eq:sepd} allows one to ask.

Two further limits deserve emphasis. First, the circuit-evaluation budget of the \avqkan\ is large (Table~\ref{tab:resources}): 25{,}000 to 110{,}000 statevector preparations per training run, compared with 10{,}000 for the quantum neural network, because the derivative-free optimiser is restarted at every step and the growth rule scans the whole pool. Under exact simulation these are circuit executions, not measurement shots, and the finite-shot runs show the model tolerating 256 shots per evaluation; the total shot count on hardware would nevertheless be of order $10^7$ per training run. Second, the model does not learn a two-dimensional classification task under a binary cross-entropy cost (Appendix~\ref{app:cls}): its error rate equals that of majority voting, while a quantum neural network and a support vector machine reach 15\%. The single-string readout with two encoded qubits does not expose a decision boundary that the growth rule can find.

\section{Conclusion}\label{sec:conclusion}

We have evaluated an adaptively grown variational quantum Kolmogorov--Arnold network under a seed-paired, leak-free, and partly pre-registered protocol, with tuned quantum and regularised classical baselines. Two findings follow. The model with 32 parameters generalises better than unregularised classical regressors and a tuned quantum neural network when ten training points are spread over 12 or 16 dimensions, with Holm-corrected $p\le0.017$, the result survives 256-shot measurement noise and executes on IBM hardware at four and twelve qubits with no measurable degradation; in the same regime a kernel ridge regressor with cross-validated hyperparameters reaches the same accuracy, and the quantum model's error is nearly independent of the number of training points, which identifies the effect as the implicit regularisation of a low-capacity model. Published quantum-KAN comparisons~\cite{Wakaura_VQKAN_2024,Wakaura2025EVQKAN} reported advantages over quantum neural networks at four qubits; under paired seeds and training-cost model selection, that difference is absent.

The reference point established here is a small one, but it is reproducible: every number in this paper traces to a seeded script and a committed data file. Without such a reference, quantum-KAN proposals will continue to be judged against untuned baselines and single runs, and claims of advantage will continue to be produced by chance. The next step is not a larger simulation but a target family on which the Kolmogorov--Arnold structure of the quantum model is expected, on theoretical grounds, to beat the RBF kernel; the standardised family and the protocol released here are designed so that such a family can be tested at the cost of a few hours of computation.

\section*{Data and code availability}
The reference implementation, all experiment drivers, the pre-registration document, the frozen analysis script, and every per-step result file underlying the figures and tables are available at \url{https://github.com/WakauraH/avqkan-model} (commit history preserved; the pre-registration precedes the confirmatory runs). The hardware results are stored with their IBM Quantum job identifiers (\texttt{dagepnb9k43c73ad9390}, \texttt{dageqs39k43c73ad9630}).

\section*{Acknowledgements}
The authors thank the developers of the blueqat SDK.

\bibliographystyle{quantum}
\bibliography{refs}

\appendix

\section{Classification}\label{app:cls}

To test the model outside regression we use a two-dimensional classification task: points $\mathbf{x}\in[0,1]^2$ are labelled $+1$ if $u_1>\tilde g(u_0)$ and $-1$ otherwise, with $u=2\mathbf{x}-1$ and $\tilde g$ the boundary $g(u_0)=e^{d_0u_0+d_1}+d_2\sqrt{1-d_3u_0^2}+\cos(d_4u_0+d_5)+\sin(d_6u_0+d_7)$, with coefficients $d_k\sim U[0,1]$ drawn per seed, min--max normalised to $[-1,1]$ so that the classes are balanced. The \avqkan\ and the quantum neural networks are trained with the binary cross-entropy of $p=(h+1)/2$, all other settings as in Sec.~\ref{sec:model}, and evaluated by the number of misclassified test points out of 50 at the step of minimum training cost. Over ten seeds the \avqkan\ misclassifies $22.9\pm6.8$ points with the cost-based growth rule and $23.9\pm6.0$ with the gradient-based rule, indistinguishable from majority voting ($24.5\pm7.3$, $p=0.75$). The three-layer quantum neural network misclassifies $7.7\pm3.1$ points, the tuned network $7.4\pm3.4$, and a support vector machine with an RBF kernel $7.7\pm4.1$ ($p=0.002$ against the \avqkan\ in each case); logistic regression reaches $18.8\pm10.7$. The growth rule adopts operators on this task, so the failure is not one of optimisation stalling but of the readout $\braket{Z_0Z_1}$ of a two-qubit-encoded state offering no sign change that the spline-modulated rotations can place along the boundary.

\section{Full pre-registered analysis}\label{app:confirm}

Table~\ref{tab:confirm} reproduces the output of the frozen analysis script for the four pre-registered hypotheses and lists the exploratory comparisons against each classical model.

\begin{table*}[t]
\centering
\caption{Pre-registered hypotheses (seeds 30--49, $n=20$, Holm $m=4$) and exploratory per-model comparisons (Benjamini--Hochberg $q$). Errors are means $\pm$ s.d.; differences are median paired differences (negative favours the \avqkan).}
\label{tab:confirm}
\small
\begin{tabular}{llcccc}
\toprule
hypothesis & comparator & \avqkan & comparator & median diff. & $p$ (adj.) \\
\midrule
H12 & classical oracle & $11.14\pm1.95$ & $13.03\pm2.48$ & $-2.44$ & 0.017 (Holm) \\
H16 & classical oracle & $11.74\pm3.13$ & $15.01\pm2.00$ & $-3.95$ & 0.004 (Holm) \\
H18 & classical oracle & $16.28\pm3.53$ & $15.79\pm1.92$ & $-0.01$ & 1.00 (Holm) \\
H12, 256 shots & classical oracle & $10.80\pm2.46$ & $13.03\pm2.48$ & $-3.01$ & 0.013 (Holm) \\
\midrule
H12 & classical KAN & & $14.18\pm2.06$ & $-2.86$ & $<10^{-4}$ (BH) \\
H12 & MLP $8\times8$ & & $17.35\pm3.05$ & $-5.73$ & $<10^{-4}$ (BH) \\
H12 & MLP 16 & & $17.24\pm5.19$ & $-5.63$ & $10^{-4}$ (BH) \\
H12 & linear & & $18.38\pm3.69$ & $-7.66$ & $<10^{-4}$ (BH) \\
H12 & tuned QNN & & $13.90\pm1.66$ & $-2.40$ & $<10^{-4}$ (BH) \\
H16 & classical KAN & & $15.11\pm1.93$ & $-3.95$ & 0.001 (BH) \\
H18 & classical KAN & & $15.89\pm1.92$ & $-0.19$ & 0.90 (BH) \\
\midrule
H12 & kernel ridge (post hoc) & & $11.15\pm3.52$ & $+0.58$ & 0.96 \\
H16 & kernel ridge (post hoc) & & $13.70\pm3.18$ & $-2.94$ & 0.048 \\
H18 & kernel ridge (post hoc) & & $14.43\pm2.73$ & $+0.52$ & 0.23 \\
\bottomrule
\end{tabular}
\end{table*}

\end{document}